\documentclass[preprint,showpacs,preprintnumbers,amsmath,amssymb]{revtex4}

\usepackage{graphicx}
\usepackage{dcolumn}
\usepackage{bm}

\begin{document}




\title{Stable and polarized Betatron x-ray radiation from a laser plasma accelerator in ionization injection regime}

\author {Andreas Doepp$^{1,2}$, Benoit Mahieu$^{1}$, Antoine Doche$^{1}$, Cedric Thaury$^{1}$, Emilien Guillaume$^{1}$, Agustin Lifschitz$^{1}$, Gabriele Grittani$^{3}$,  Olle Lund$^{4}$, Martin Hansson$^{4}$, Julien Gautier$^{1}$, Michaela Kozlova$^{4}$, Jean Philippe Goddet$^{1}$, Pascal Rousseau$^{1}$, Amar Tafzi$^{1}$, Victor Malka$^{1}$, Antoine Rousse$^{1}$, Sebastien Corde$^{1}$ and Kim Ta Phuoc$^{1}$}

\affiliation{$^{1}$ Laboratoire d'Optique Appliqu\'ee, ENSTA, CNRS UMR7639, Ecole Polytechnique, Chemin de la Huni\`ere,91761 Palaiseau, France.}

\affiliation{$^{2}$ Centro de Laseres Pulsados, Parque Cientfico, 37185 Villamayor, Salamanca, Spain.} 

\affiliation{$^{3}$ ELI Beamlines Project, Institute of Physics of the ASCR, Na Slovance 2, 182 21 Prague 8, Czech Republic.}

\affiliation{$^{4}$ Department of Physics, Lund University
PO Box 118, 22100 Lund, Sweden.}

\begin{abstract}

Betatron x-ray source from laser plasma interaction combines high brightness, few femtosecond duration and broad band energy spectrum. However, despite these unique features the Betatron source has a crippling drawback preventing its use for applications. Its properties significantly vary shot-to-shot and none of the developments performed so far resolved this problem. In this letter we present a simple method that allows to produce stable and bright Betatron x-ray beams. In addition, we demonstrate that this scheme provides polarized and easily tunable radiation. Experimental results show that the pointing stability is better than 10$\%$ of the beam divergence, with flux fluctuation of the order of 20$\%$ and a polarization degree reaching up to 80$\%$. 

\end{abstract}

\pacs{52.38.Ph,52.25.Os,52.38.-r,52.50.Dg}

\maketitle

\medskip
\medskip

Laser-produced Betatron radiation is a femtosecond x-ray beam emitted by relativistic electrons in a laser-plasma accelerator \cite{RoussePRL2004,JoshiPRL2002}. This all optical x-ray source reproduces in a millimeter scale the principle of conventional synchrotrons \cite{CordeRMP2013}. Here, an ion cavity driven by an intense laser pulse acts as both an electron accelerator and a wiggler. Demonstrated ten years ago this source can deliver x-ray beams in the few kiloelectronvolt energy range with brightnesses similar to third generation synchrotrons \cite{KneipNatPhys2010}. However, while this source could open a wide range of novel applications in multidisciplinary fields, its use is limited to basic applications because of significant shot to shot fluctuations of essential properties such as flux, spectrum, pointing and spatial profile. In addition, the source polarization can not be reliably controlled. Nonetheless, these issues can in principle be solved by controlling orbits of relativistic electrons emitting the radiation \cite{CordeRMP2013}.

The most efficient laser-plasma interaction regime to produce Betatron radiation is to date the so called bubble regime \cite{PukhovAPB2002}. In this case the ponderomotive force of an intense femtosecond laser pulse  propagating in an underdense plasma pushes electrons away from high intensity regions and drives, in its wake, a relativistic plasma wave. The first period of this plasma wave consists in an ion cavity almost free of background electrons. Electrons injected into this cavity are accelerated to hundreds of megaelectronvolt energies in the laser propagation direction \cite{MalkaScience2002}, wiggled in the transverse direction with a period of a few hundreds microns, and emit Betatron x-ray radiation.  
The properties of Betatron radiation depend exclusively on the electron orbits in the cavity \cite{CordeRMP2013}. As such, the position and momentum of electrons at injection determines many radiation features. In most experiments, electrons injection into the wake relies on transverse self-injection where electrons travel along the bubble sheath and are injected at the back of the cavity \cite{CordeNatCom2013}. In this scheme the electron positions at injection cannot be easily controlled because it strongly depends on the evolution of the laser as it propagates in the plasma. Consequently, the properties of the x-ray radiation strongly vary shot to shot. In principle, the x-ray beam features may be stabilized and controlled using  injection techniques such as colliding injection \cite{FaureNature2006} or density transition injection \cite{SchmidPSTAB2010} which have been used to demonstrate the production of stable and controllable electron beam. However, the beam charge and the transverse amplitude of the electron motion are too weak to efficiently produce Betatron radiation \cite{CordePRL2011}.

In this letter we present Betatron radiation produced in a regime, based on the injection of electrons from tunnel ionization \cite{PakPRL2010,McGuffeyPRL2010,ChenPOP2012}, that allows to overcome these limitations. This process relies on the use of a plasma from a gas mixture made of a low atomic number gas (here Helium) with a small amount of a high atomic number gas (here 1$\%$ of Nitrogen). In our interaction regime, the front edge of the laser pulse propagating in the mixture fully ionizes Helium and the five outer electrons of Nitrogen. The ponderomotive force pushes electrons and creates the wakefield cavity. At the vicinity of the  peak laser intensity, electrons are released by ionization of N$^{5+}$ and N$^{6+}$. Born at rest into the cavity, these electrons can gain a sufficient longitudinal energy from the wakefield to be trapped, accelerated and wiggled. 

This injection mechanism is a particular type of longitudinal injection \cite{CordeNatCom2013}. As such, it is less sensitive to the transverse inhomogeneities of the laser and to laser propagation than transverse injection, and it leads to the production of stable electrons beams. In addition, electrons gain a net transverse momentum from the laser field along its polarization direction. They therefore have a preferential plane of oscillation and may have a wider transverse distribution than in the case of transverse self-injection \cite{ChenPOP2012}. These features are promising to produce bright, stable, and polarized Betatron radiation. First, initial conditions of the electrons orbits are well defined. Second, a wider transverse amplitude of motion leads to the production of higher flux and more energetic radiation. Finally, electrons orbits being essentially confined in the plane of the laser polarization axis, the radiation is expected to be polarized in this plane and tunable. 

We performed 3D Particle In Cell simulations. In this simulation, we consider a 30 fs full width half maximum (FWHM) laser pulse focussed in a 14 $\mu$m FWHM waist. The normalized vector potential is $a_0=1.2$ and the electron plasma density is $1.5 \times 10^{19}$ cm$^{-3}$. The gas is a mixture of He (99$\%$) and Nitrogen (1$\%$). The density profile consists in a 1.2 mm ascending ramp, a 300 $\mu$m plateau and a 1.2 mm descending ramp. Most of injected electrons originate from ionization of N$^{5+}$ and N$^{6+}$. Figure 1 shows typical transverse electron orbits. Electrons are released close to the center of the cavity, where the laser intensity is maximum. The orbit of an electron depends on its transverse position at ionization with respect to the vertical axis of the cavity (here aligned with the laser polarization axis). If electrons are initially close to the vertical axis, the laser field and the radial field of the cavity are aligned. Electrons gain momentum along the polarization axis and their motion is mainly confined in this plane. If electrons are ionized further from the vertical axis, the radial cavity field and laser field are not aligned anymore. Electrons have elliptical orbit whose axis makes an angle with the polarization plane. The transverse amplitude of motion can be much smaller in that case, as shown in Figure 1c. Nevertheless, the numerical simulation shows that a majority of the electrons will preferably oscillate along the direction of the laser polarization and with a transverse amplitude of a few microns.

We experimentally characterized the Betatron radiation produced in this tunnel ionization injection regime. The experiment was performed at the Laboratoire d'Optique Appliqu\'ee using a Titanium-doped Sapphire (Ti:Sa) laser operating at 1 Hz with a wavelength $\lambda _0$ of 800 nm. The laser delivered energies up to 1.2 Joule on target. The pulse duration is 28 fs FWHM. The laser beam was focused with an f/10 off-axis parabolic mirror onto the edge of a 3 mm supersonic gas jet.  The laser distribution in the focal plane was Gaussian with a waist $w_0$ of 18 $\mu m$ FWHM. This produces vacuum-focused intensities $I_L$ on the order of $3.5 \times 10^{18}$ W/cm$^2$, for which the corresponding normalized vector potential $a_0$ is 1.2. The linear laser polarization was adjusted using a half-wave plate. We measured the electron energies using a permanent magnet (0.7 T over 40 cm) deviating electrons onto a phosphor scintillator. The plasma electron density was $1\times 10^{19}$ cm$^{-3}$.

In this parameter regime, electrons from the laser plasma accelerator had a broadband spectrum extending up to about 250 MeV. The divergence of the electron beam was about $\theta_\parallel = 16$ mrad FWHM and $\theta_\perp = 4$ mrad FWHM along the direction parallel and perpendicular to the laser polarization axis. Note that this asymmetry in the electron beam profile is also in agreement with source size measurement (it is 5$\pm$ 1 $\mu$m along the laser polarization and 1.7 $\pm$ 1 $\mu$m in the perpendicular direction). The charge contained in the bunch was about 40 pC. X-ray radiation  was measured using either an indirect detection x-ray CCD camera (Princeton QuadRO) placed on the laser axis at 70 cm from the source or a direct detection x-ray camera (Princeton Pixis) placed on the laser axis at 9 m from the source. For all measurements, a 50 $\mu$m thick Mylar window was kept in front of the CCD camera. Indirect detection camera was used for the angular profile measurements (in that case an additional 300 $\mu$m thick Berylium filter is placed in front of the camera). Direct detection camera was used for single photon counting spectrum measurements (in that case additional 26 $\mu$m thick Aluminium filter and 50 $\mu$m thick Mylar were placed in front of the camera).

We first studied the stability of the x-ray beam profile in the gaz mixture with respect to pure Helium. Figure 2a shows four consecutive shots in the gas mixture and in pure Helium. In pure Helium, the shape of the x-ray beam fluctuates shot to shot in an uncontrolled way \cite{TaPhuocPRL2006}. Round, elliptical and annular beam profiles were typically observed. Note that x-ray beams profiles are signatures of electrons orbits in the cavity \cite{TaPhuocPRL2006}. Therefore, fluctuations of the x-ray angular profile generally result in fluctuations of flux and spectrum of the radiation. In contrast, in a gas mixture, the beam has consistently the same elliptical profile and is highly stable. As an illustration, Figure 2b represents the centroid position of fifty consecutive shots and the sum of the fifty beam profiles. The averaged beam has an elliptical shape with FWHM divergences $\theta_X=33$ mrad and $\theta_Y=12$ mrad which is very close to the divergences of each individual beam. The standard deviation of the centroid position is 1 mrad, which is about 10$\%$ of the beam divergence. 
Stable x-ray beam profiles result in stable x-ray beam properties (flux and spectrum). Using the same set of images we studied the stability of the maximum x-ray flux. The result is shown in Figure 3a. The standard deviation of the maximum signal is about 15$\%$. We then studied the stability of the radiation spectrum. For this measurement, we placed the direct detection camera at 9 meters from the source and we use single photon counting method \cite{FourmauxNJP2011}. In our configuration we could measure the spectrum from 4 to 14 keV. Figure 3a represents a typical spectrum and a fit with a synchrotron spectrum. The critical energy measured is presented in figure 3b for sixty consecutive shots. The mean critical energy is $6.7$ keV and the standard deviation is 0.46 keV. In pure Helium, the mean critical energy is slightly lower ($E_c=6.5$ keV) and the standard deviation is twice larger.

The x-ray beam profile being a direct signature of the electron orbits and therefore of the emitted electromagnetic field, its measurement provides information on the polarization of the radiation \cite{TaPhuocPRL2006}. In pure Helium experiments the radiation can be polarized on some shots  \cite{SchnellNatCom2013}. However, none of the injection methods used so far allow to reliably control the polarization of the x-ray beam. In colliding injection, as in the bubble regime, the electron distribution at injection can not be controlled. Here, in contrast, the beam shape and hence the x-ray beam polarization are very stable because electrons oscillate in a preferential direction \cite{Jackson}. Figure 4 shows x-ray beam profiles recorded for four laser polarization directions. For all shots, the angular profile has an elliptical shape. The major ellipse axis is always aligned along the laser polarization axis. The divergence of the x-ray beam is 24 $\pm$ 1 mrad FWHM and 10 $\pm$ 1 mrad FWHM along the major and minor axis respectively. Note that the beam profile had a circular shape for circular laser polarization (not shown). The angular distribution measured indicates that the x-ray radiation follows the polarization of the laser field. In addition the fact that the angular shape of the x-ray beam remains identical for all polarization indicates that injection is symmetrical around the laser axis. Indeed, in the case of asymmetric injection we would have observed different shapes for the angular profile of the x-ray beam when varying the orientation of the half wave plate.

We performed test particule simulations to reproduce x-ray beam profiles observed experimentally and estimate the polarization degree of the radiation. We assume electrons born at rest at the maximum intensity of the laser pulse. Conservation of canonical momentum implies that the transverse momentum gain from the laser field along the polarization direction ranges from zero to  $p_\parallel = a_0$ in the plasma \cite{MoorePRL1999}. We assume the wake to be a spherical bubble, electrons initially distributed in a 1.2 micron diameter disk and calculate the electron orbits as described in reference \cite{TaPhuocPoP2008}. The properties of the radiation is then obtained by integrating the general expression of the radiation emitted by a relativistic electron in motion \cite{Jackson}. The FWHM contour of radiation angular profile obtained by the numerical simulation is presented in Figure 4b. For this simulation, we assume a uniform distribution of $p_\parallel $ from 0 to 3 and $n_e=1.0\times 10^{19}$ cm$^{-3}$.  From this simulation we can estimate the polarization degree $D_p$. It is defined as the percentage of intensity radiated along the laser polarization axis with respect to the total emitted intensity. The intensity of the radiation emitted at a given polarization is obtained by taking the scalar product of the vectorial integral in the general expression of the radiation emitted by a moving charge \cite{Jackson} by the polarization direction. Here we obtain a polarization degree of $D_p = 80 \pm 5 \%$. 


In conclusion, we reported on Betatron radiation from a laser plasma accelerator in the tunnel ionization injection regime. This interaction regime allows to produce x-ray beams in the keV range with unprecedented stability (pointing, flux and spectrum) and controlled polarization. For a given polarization, 100$\%$ of the x-ray beams had the same polarization state. Future developments will focus on further increasing the polarization degree of the radiation, for example, by using a mixture of Helium and Neon to ionize the electrons closer to the peak intensity region of the laser field. In addition, the beam profile measurement as a function of the laser polarization can become a tool to estimate the electron distribution at injection. Indeed, asymmetric injection would result in profiles elliptical along the laser polarization and more circular in the transverse direction. We also anticipate that stable and polarized Betatron radiation will open novel possibilities for application experiments in multidisciplinary fields of ultrafast x-ray science.\\

ACKNOWLEDGMENTS: We acknowledge the Agence Nationale pour la Recherche through the FENICS Project No. ANR-12-JS04-0004-01, the Agence Nationale pour la Recherche through the FEMTOMAT Project No. ANR-13-BS04-0002, LA3NET project (GA-ITN-2011-289191), and GARC project 15-03118S.

\medskip
\medskip

\clearpage

\begin{figure}
\includegraphics[width=10cm]{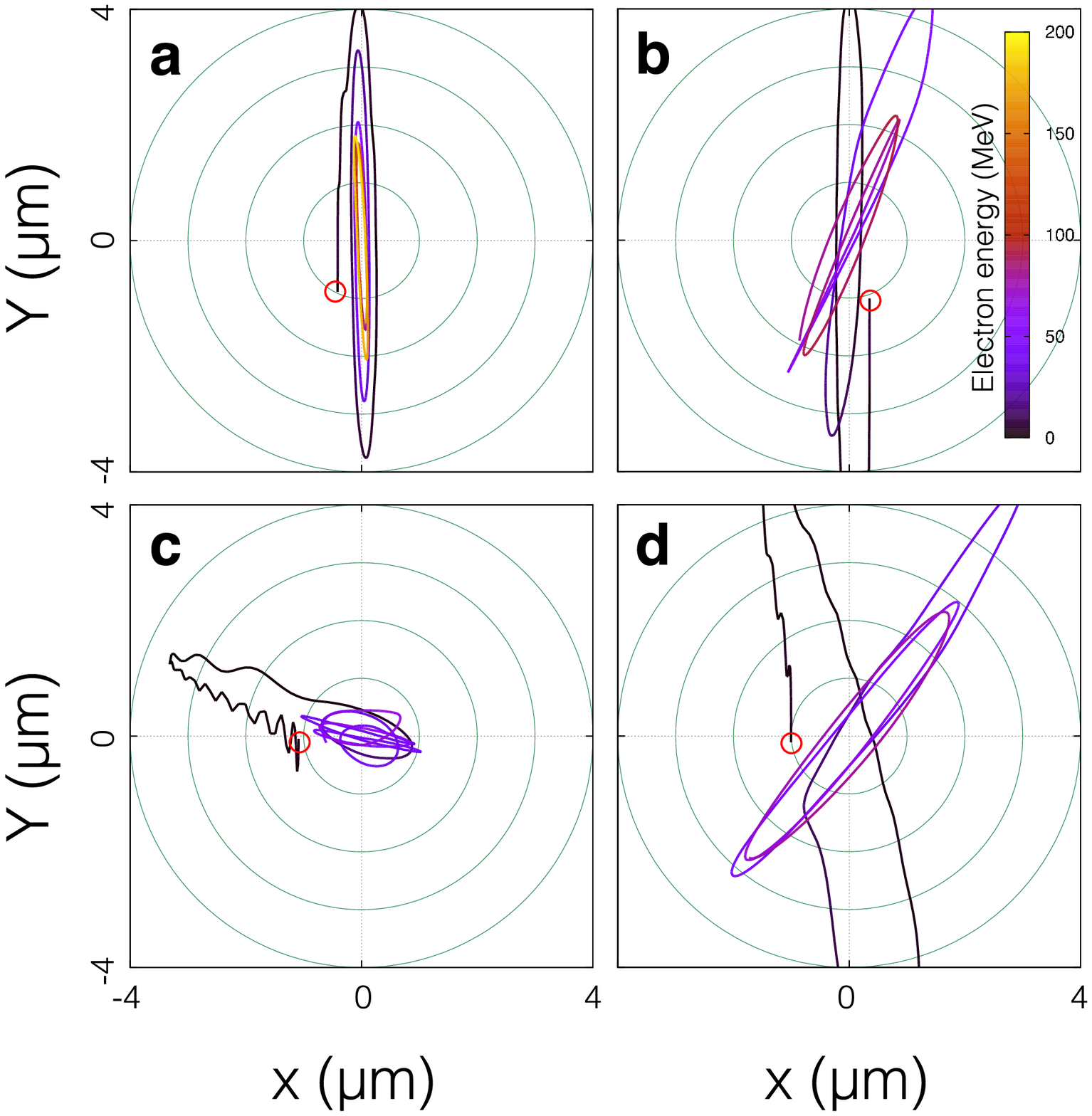}
\caption{Transverse electron orbits from 3D PIC simulations. The results show that the electron trajectory depends on its position at ionization. 1a) Electron is initially close to the vertical axis of the cavity. Electron gains momentum along the polarization axis and its motion is mainly confined in this plane. 1c) Electron is ionized further from the vertical axis. The radial cavity field and laser field are not aligned anymore, and the orbit is elliptical. The transverse amplitude of motion can be much smaller in that case.}
\end{figure}

\begin{figure}
\includegraphics[width=10cm,angle =180]{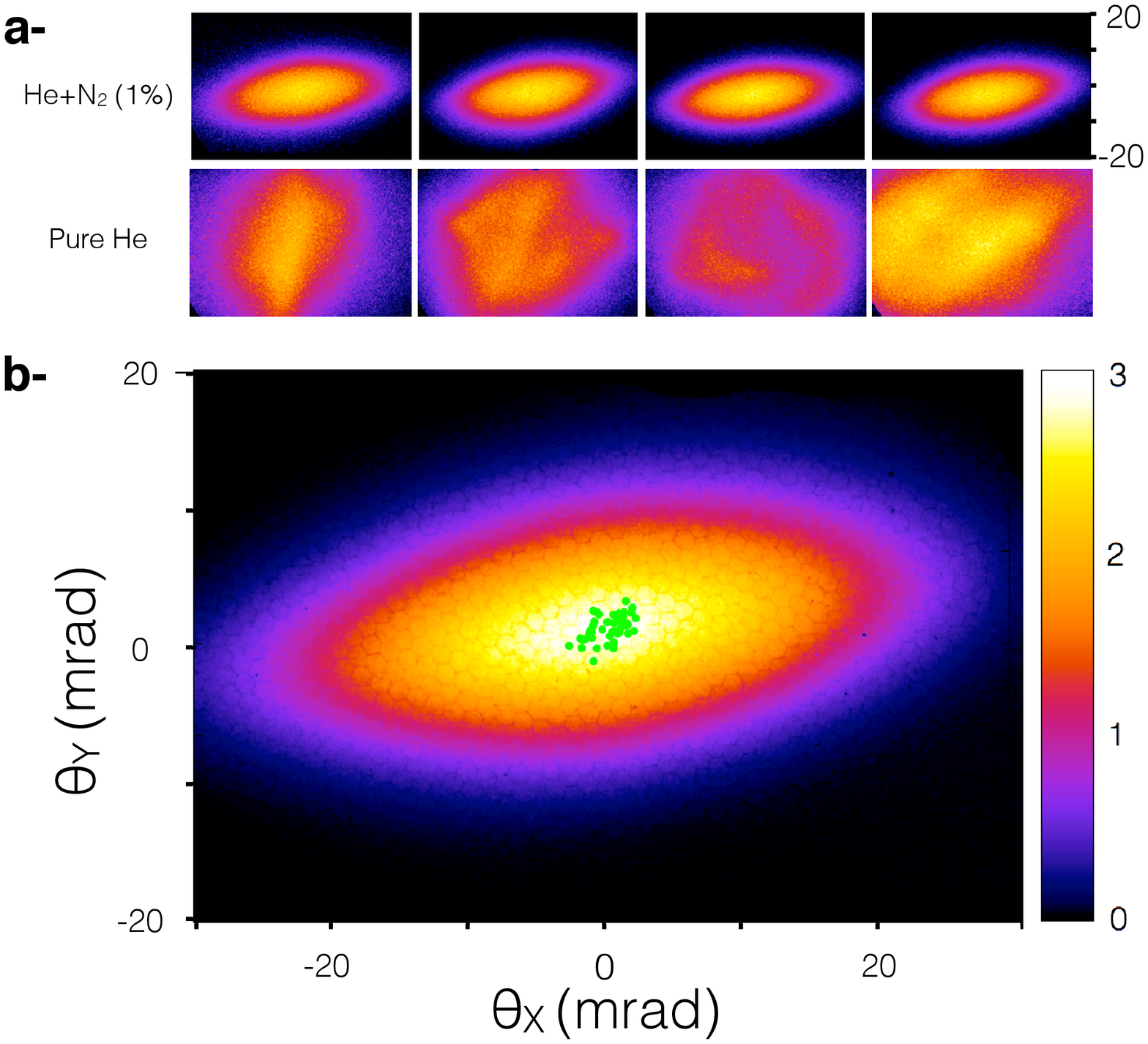}
\caption{a- Angular profile of the x-ray beam for four consecutive shots in the gaz mixture (He+N$_2$ 1$\%$) and in pure Helium. The color scale is the same for all images. b- Position of the centroid position of the x-ray beam for 50 consecutive shots. Each green dot represent the centroid position of one x-ray beam. The standard deviation is 1 mrad, which corresponds to about 10$\%$ of the beam FWHM divergence. The beam profile shown is the sum of the 50 shots. The elliptical beam profile is highly reproducible. The FWHM divergences are 33 mrad and 12 mrad along the two axis of the ellipse.}
\end{figure}

\begin{figure}
\includegraphics[width=10cm]{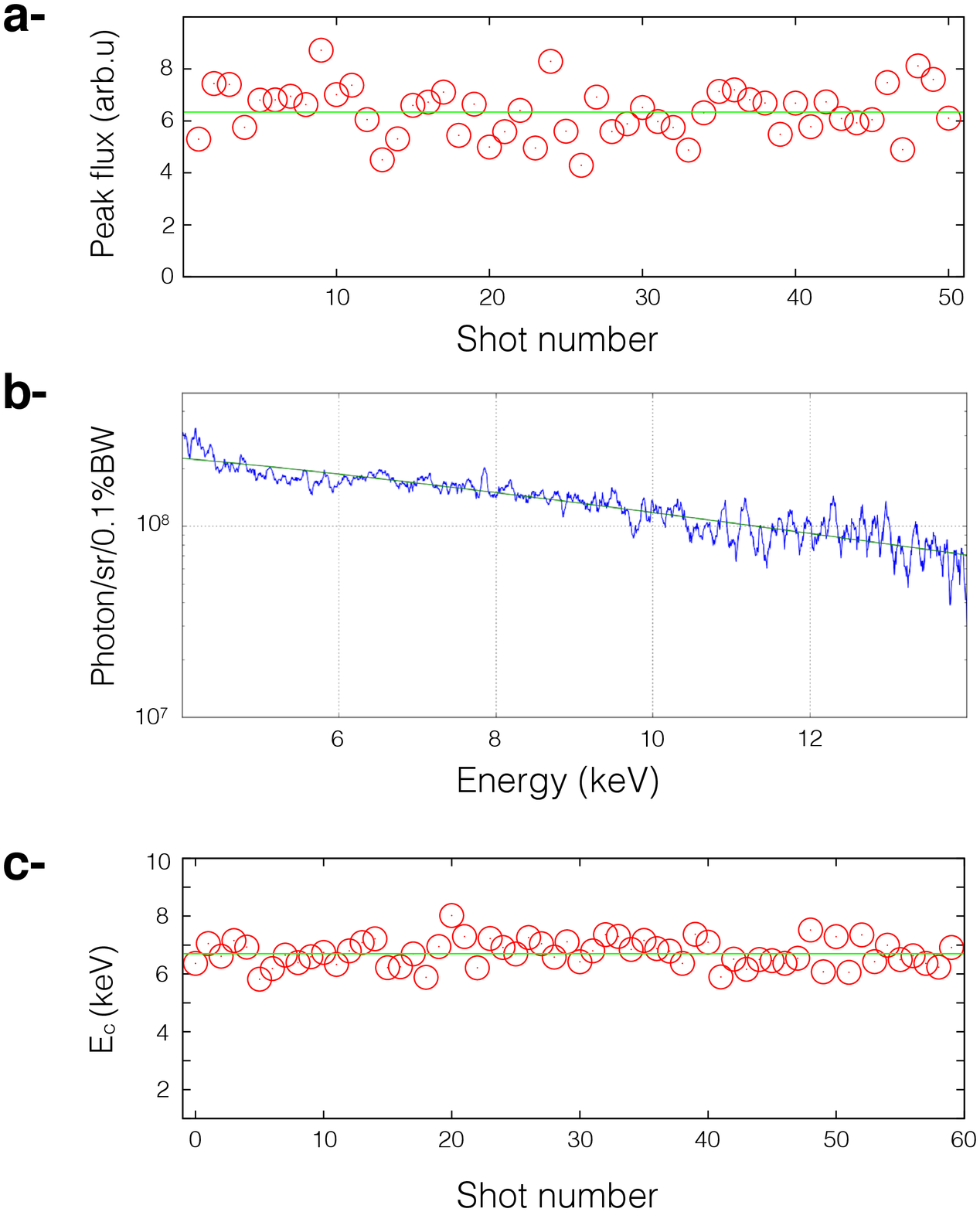}
\caption{a- Flux fluctuation measured for fifty consecutive shots. Each red dot represents the maximum flux measured for one shot. The green line is the averaged flux for the fifty shots. The standard deviation of the maximum flux is about 20$\%$ of its mean value. b- Typical spectrum measured by single photon counting. The data is fitted using a synchrotron function. c- Critical energy for sixty consecutive shots. Each dot represents the critical energy for one shot.  The green line is the averaged critical energy for the sixty shots.}
\end{figure}

\begin{figure}
\includegraphics[width=10cm, angle =-90]{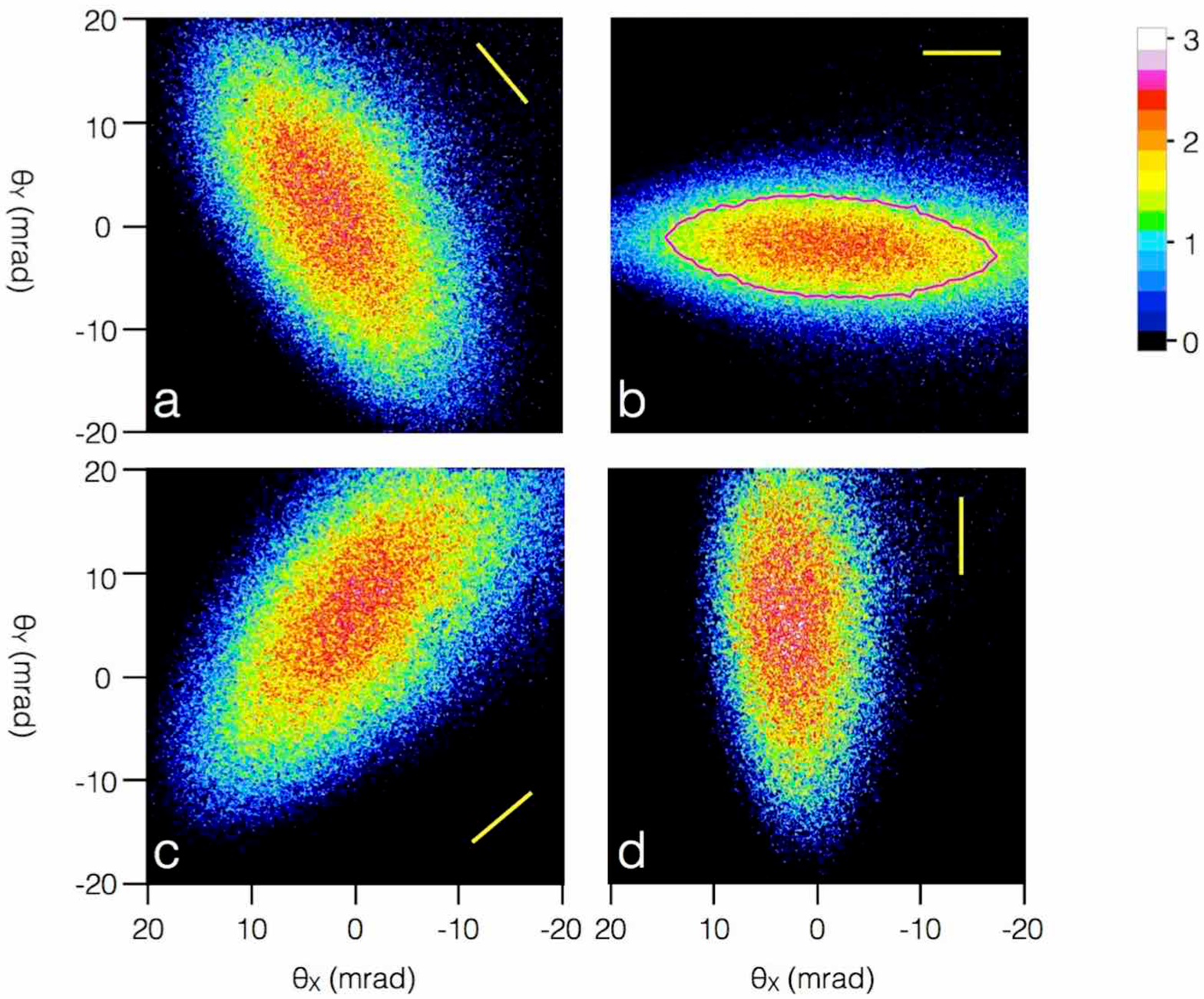}
\caption{Betatron x-ray beam profiles measured for four orientation of the laser polarization. The laser polarization axis is indicated by the yellow line. The line in figure 4b represents the FWHM contour of the beam profile obtained from the test particule simulation.}
\end{figure}


\begin{references}

\bibitem{RoussePRL2004}
Rousse A., Ta Phuoc K., Shah R., Pukhov A., Lefebvre E., Malka V., Kiselev S. et al., Phys. Rev. Lett., 93(13), 135005 (2004).

\bibitem{JoshiPRL2002}
Wang S, Clayton C.E, Blue B.E, Dodd E.S, Marsh K.A, et al., Phys. Rev. Lett., 88(13), 135004 (2002).

\bibitem{CordeRMP2013}
Corde S., Ta Phuoc K., Lambert G., Fitour R., Malka V., Rousse A., Beck A., et al., Reviews of Modern Physics, 85(1), 1-48 (2013).

\bibitem{KneipNatPhys2010}
Kneip S., McGuffey C., Martins J. L., Martins S. F., Bellei C., Chvykov V., Dollar F., et al., Nat. Phys., 6(12) (2010).

\bibitem{PukhovAPB2002}
Pukhov, A. and Meyer-ter-Vehn, J., Applied Physics B: Lasers and Optics, 74(4-5), 355-361 (2002).

\bibitem{MalkaScience2002}
Malka V., Fritzler S., Lefebvre E., Aleonard M-M., Burgy F., Chambaret J-P., Chemin J.F, et al., 298, 1596 (2002).

\bibitem{CordeNatCom2013}
Corde S., Thaury C., Lifschitz A., Lambert G., Ta Phuoc K., Davoine X., Lehe R., Douillet D., Rousse A., and Malka V., Nat Commun 4, 1501 (2013).

\bibitem{SchmidPSTAB2010}
Schmid K., Buck A., Sears C.M.S, Mikhailova J.M, Tautz R., Herrmann D., Geissler M., Krausz F., and Veisz L., Phys. Rev. STAB, 13, 091301 (2010)

\bibitem{FaureNature2006}
Faure J., Rechatin C., Norlin A., Lifschitz, A. Glinec Y., Malka V., Nature 444, 737 (2006). 

\bibitem{CordePRL2011}
S. Corde, K. Ta Phuoc, R. Fitour, J. Faure, A. Tafzi, J. P. Goddet, V. Malka, and A. Rousse, Phys. Rev. Lett. 107, 255003 (2011).

\bibitem{PakPRL2010}
Pak, A., Marsh, K. A., Martins, S. F., Lu, W., Mori, W. B., Joshi, C., Physical Review Letters, 104(2), 025003 (2010).

\bibitem{McGuffeyPRL2010}
C. McGuffey, A. G. R. Thomas, W. Schumaker, T. Matsuoka, V. Chvykov, F. J. Dollar, G. Kalintchenko, V. Yanovsky, A. Maksimchuk, K. Krushelnick, V. Y. By- chenkov, I. V. Glazyrin, and A. V. Karpeev, Physical Review Letters 104, 025004 (2010).

\bibitem{ChenPOP2012}
Chen M., Esarey E., Schroeder C. B., Geddes C. G. R., and Leemans W. P., Physics of Plasmas, 19, 033101 (2012) .

\bibitem{TaPhuocPRL2006}
Ta Phuoc K., Corde S., Shah R., Albert F., Fitour R., Rousseau J.-P., Burgy F., et al., Phys. Rev. Lett., 97(22), 225002 (2006).

\bibitem{FourmauxNJP2011}
Fourmaux S., Corde S. Ta Phuoc K., Leguay P-M., Payeur S., Lassonde P., Gnedyuk S., Lebrun G., Fourment C, Malka V, Sebban S, Rousse A and Kieffer J-C., New Journal of Physics, 13, 033017 (2011).  

\bibitem{SchnellNatCom2013}
Schnell, M., Sa, A., Jansen, O., Pukhov, A., Kaluza, M. C., Ja, O., Spielmann, C. Nature Communications 4, 2421 (2013).

\bibitem{MoorePRL1999}
Moore C.I, Ting A., McNaught S. J. , Qiu J., Burris H. R., Sprangle P., Phys. Rev. Lett., 82(8), 1688 (1999).

\bibitem{TaPhuocPoP2008}
Ta Phuoc K, Esarey E., Leurent V., Cormier-Michel E., Geddes C.G.R, Schroeder C.B., Rousse A. and Leemans W.P., Physics of Plasmas 15, 063102 (2008).

\bibitem{Jackson}
J. D. Jackson, Classical electrodynamics, 3rd edition, {\it Classical electrodynamics} (Wiley, New York 2001).



\end{references}
\end{document}